\documentclass[aip,jcp,reprint,floatfix,superscriptaddress]{revtex4-1}
\usepackage{amsfonts}
\usepackage{amsmath}
\usepackage{graphicx}

\newcommand{\comment}[1]{}

\newcommand{\lbrs}{\left[}
\newcommand{\rbrs}{\right]}

\newcommand{\beq}{\begin{eqnarray}}
\newcommand{\eeq}{\end{eqnarray}}
\renewcommand{\d}{{\rm{d}}}

\newcommand{\half}{\frac{1}{2}}
\newcommand{\aeq}{\approx}

\newcommand{\Liue}{Ref.~\onlinecite{Liu2012}}

\newcommand{\EB}{E_{\textrm{BLB}}}
\newcommand{\ECl}{E_{\textrm{Clv}}}

\newcommand{\sus}[1]{${}_{\textrm{#1}}$}
\newcommand{\sps}[1]{${}^{\textrm{#1}}$}

\newcommand{\PhiLJ}{\Phi_{\rm{LJ}}}
\newcommand{\PhiNCM}{\Phi_{\rm{NCM}}}
\newcommand{\PhiLJp}{\Phi_{\rm{LJ}(p)}}

\newcommand{\Jmm}{J/m\sps{2}}

\usepackage[usenames,dvipsnames]{color}
\newcommand{\changed}{{}}

\begin{document}
\title{Binding and interlayer force in the near-contact region of
two graphite slabs: experiment and theory}
\author{Tim Gould}
\email{t.gould@griffith.edu.au}
\affiliation{Queensland Micro and Nano Technology Centre, Nathan campus, %
Griffith University, 170 Kessels Road, Nathan, QLD 4111, Australia}
\author{Ze Liu}
\affiliation{Department of Engineering Mechanics and Center for %
Nano and Micro Mechanics, Tsinghua University, Beijing 100084, China}
\author{Jefferson Zhe Liu}
\email{zhe.liu@monash.edu}
\affiliation{Department of Mechanical and Aerospace Engineering, %
Monash University, Clayton, VIC 3800, Australia}
\author{John F. Dobson}
\affiliation{Queensland Micro and Nano Technology Centre, Nathan campus, %
Griffith University, 170 Kessels Road, Nathan, QLD 4111, Australia}
\author{Quanshui Zheng}
\affiliation{Department of Engineering Mechanics and Center for %
Nano and Micro Mechanics, Tsinghua University, Beijing 100084, China}
\affiliation{Institute of Advanced Study, %
Nanchang University, Nanchang, China}
\author{S. Leb\`egue}
\affiliation{Laboratoire de Cristallographie, R\'esonance Magn\'etique %
et Mod\'elisations (CRM2, UMR CNRS 7036) Institut Jean Barriol,
Universit\'e de Lorraine BP 239,
Boulevard des Aiguillettes 54506 Vandoeuvre-l\`es-Nancy, France}

\begin{abstract}
Via a novel experiment, Liu \emph{et al.}
[Phys. Rev. B, {\bf 85}, 205418 (2012)]  estimated the graphite binding
energy, specifically the cleavage energy, an important physical
property of bulk graphite. We re-examine the data
analysis and note that within the standard Lennard-Jones model
employed, there are difficulties in achieving internal consistency
in the reproduction of the graphite elastic properties.
By employing similar models which guarantee consistency with the
elastic constant, we find a wide range of model dependent binding
energy values from the same experimental data.
We attribute some of the difficulty in the determination of the
binding energy to: i) limited theoretical understanding of the
van der Waals dispersion of graphite cleavage,
ii) the mis-match between the strong bending stiffness of the
graphite-SiO$_2$
cantilever and the weak asymptotic inter-layer forces that are
integrated over to produce the binding energy.
We find, however, that the data does support determination of
a maximum inter-layer force that is relatively model independent.
We conclude that the peak force per unit area is $1.1 \pm 0.15$GPa
for cleavage, and occurs at an inter-layer spacing of $0.377\pm 0.013$nm.
\end{abstract}
\maketitle

\section{Introduction}

Graphene has attracted sustained interest in recent years because of
its unusual electronic, magnetic and mechanical
properties\cite{electric,r1,r2,r3,misha1}. Applications that
depend on mechanical properties include, 
for example, flexible touch-screens\cite{SamPat}
and graphene-coated oscillating sensor devices.
These can be based on large-scale high-quality 
flexible vapour deposited graphene sheets\cite{kim}.
To model such applications one requires a reliable knowledge of
the force and binding energy between graphene layers,
quantities that have recently been controversial both at the
theoretical and experimental level. For example, results
for the graphene layer binding energy vary by at least a
factor of two between different
experiments\cite{Girifalco1956,Benedict1998,Zacharia2004}.
An equally large spread of predictions is found amongst theoretical
analyses\cite{Dion2004,Chakarova2006,Ziambaras2007,Hasegawa2004,%
Spanu2009,Lebegue2010,Thrower2013,TSvasp}. Any fresh experimental insight
on this system is therefore important.

A recent paper\cite{Liu2012} from some of the authors used
the bulk mechanical properties of graphite to establish a value for the
inter-graphene-layer
binding energy indirectly from displacement measurements. In the
experiment, a flake of graphite with a SiO\sus{2} backing was
allowed to cantilever into a stepped graphite substrate and
atomic force microscopy (AFM)
used to measure the profile of the cantilever.
This profile was then fit to the prediction from a finite element
analysis using a binding force of cleavage vs. displacement relationship
derived from a parameterised Lennard-Jones (LJ) potential.
One parameter (the effective binding energy) was allowed to vary
in the LJ potential and a best fit was found to the measured data
predicting a binding energy of $0.19\pm0.01$\Jmm.

\changed{In this paper we shall first discuss certain problems in the
previous theoretical analysis of experiment\cite{Liu2012}.
The most significant of these is an internal inconsistency between
the effective elastic coefficient used in Hooke's law for
the contact layer (20.0GPa), and other layers (36.5GPa),
discussed in depth in Section~\ref{Sec:C33}. We will show that
this discrepancy comes from limitations of the popular
Lennard-Jones model when applied to layered materials, and that
layered materials present a particularly difficult case for
indirect measurements of energy. This difficulty is due to
a combination of poorly understood theory, and
variable length and force scales for different energetic
contributions.

We will then reanalyse the experimental data from
\Liue\ using improved binding energy models
to better estimate the energetics of cleavage.
These improved models remove the discrepancies in the
elastic coefficient by construction, and should thus be
considered more reliable in the near-contact regime.
We will show that the predicted binding energy varies greatly
between the different models and should
thus be considered unreliable.
However, the force vs distance curves
for intermediate interlayer distances show significantly less
variation. Unlike the energy, the force is not affected
by difficulties (discussed in Section~\ref{Sec:Stiffness}) arising
from the mismatch between the large bending stiffness of the
cantilever on one hand, and the weakness of the attractive
interlayer van der Waals forces on the other hand.

We will finally conclude that the previous theoretical analysis
on experimental results\cite{Liu2012} is likely to substantially
underestimate the cleavage energy value, but can be used
to make reliable predictions of the ``peak force'' (the minimum force
required to cleave the layers). The  peak force per unit area is
found to be $1.1 \pm 0.15$GPa, and occurs $43\pm 13$pm outside
the natural equilibrium interlayer spacing.}

\section{Theory and analysis}

As mentioned in the Introduction, we identified a number of
problems with the previous theoretical analysis of experimental
results\cite{Liu2012}. These problems highlight the
difficulties of measuring the
energetic properties of graphite, a material where direct
experiments are difficult and theory is not always
comprehensive or conclusive.
We outline the identified issues below, with the aim of
accounting for each in a reanalysis of the same data.

\subsection{Cleavage vs binding}
\label{sec:Cleav}

We first note that Liu \emph{et al.} analyzed the energy
of cleavage (the energy required to split a bulk along a plane)
rather than the bulk-layer binding energy as normally quoted
(the energy required to split a bulk into widely separated atomic planes).
A discussion of the difference between these energies is found
in the supplementary material of Ref.~\onlinecite{Bjorkman2012},
and from a more formal perspective in Ref.~\onlinecite{Gould2008}.
While these quantities are usually of the same order of magnitude,
they will differ whenever the interaction between second-neighbor
planes of atoms is significant, which it typically is in a
van der Waals bonded system such as graphene. We will subsequently use
$\ECl$ to refer to the cleavage energy and $\EB$ to refer to
the bulk-layer binding energy.

While high-level \emph{ab initio} data is not available for cleavage
of graphite, the binding energy of graphite
was found by Spanu~\emph{et al.}\cite{Spanu2009} with
Quantum Monte-Carlo (QMC) method to be
$56$meV/Atom=\allowbreak $0.34$\Jmm, and by
Leb\`egue~\emph{et al.}\cite{Lebegue2010}
under the Random Phase Approximation (RPA) to be
$48$meV/atom=\allowbreak $0.29$\Jmm.
Thrower~\emph{et al.}\cite{Thrower2013} used lower-level theory
benchmarked against
experimental results for adsorption of polycyclic aromatic
hydrocarbons on graphite to predict a graphite binding energy
of $57\pm 4$meV/atom=\allowbreak $0.34\pm 0.025$\Jmm.
The difference between the cleavage and binding energies has
also been found on a similar system\cite{Podeszwa2010}, and via
lower-level theory adjusted to match known high-level
theory\cite{Bjorkman2012}.

Podeszwa\cite{Podeszwa2010} extrapolated from anthracene, pyrene
and coronene to estimate the exfoliation, binding and cleavage
energies of graphite. While his model neglects long-ranged
(in the plane) plasmon interaction effects due to the finite
size of the coronenes, it is expected to yield a good estimate of
the relative strengths near contact. He finds a cleavage
energy 14\% greater than the binding energy of graphene based
on a four sheet model. This gives an absolute cleavage energy
of $\ECl=48.5$meV/Atom$=0.30$\Jmm. As Podeszwa notes, this
extrapolation neglects certain bulk binding properties of graphite,
and may underestimate the binding and cleavage energies.

In a recent publication\cite{Bjorkman2012}, Bj\"orkman~\emph{et al.}
investigated bulk properties in various two dimensional materials
using various theories. For graphite they
found a $2.5\textrm{meV}/\textrm{Ang}^2=0.04$\Jmm\ 
difference between the graphite exfoliation and graphene
binding energies. Unpublished results\cite{Bjorkman2012-PC} suggest
that there is a further 10\%-15\% increase in energy for cleavage
compared to exfoliation (an absolute value of $0.03-0.05$\Jmm\ based
on RPA and QMC results). This is of similar relative
magnitude to the energy difference found by Podeszwa\cite{Podeszwa2010},
and the energy difference predicted by LJ pair summation models
using experimental layer spacing.

These various theories thus predict a cleavage energy $\ECl$
between $0.30$\Jmm\ and $0.39$\Jmm, with a binding energy $\EB$
between $0.26$\Jmm\ and $0.34$\Jmm\ with the lower bound likely
underestimated. The \emph{cleavage} energy $\ECl=0.19\pm 0.01$\Jmm\ 
estimated via a theoretical analysis of the cantilever
experiment\cite{Liu2012} would correspond to a
binding energy of $\EB=0.17\pm 0.01$\Jmm,
around half that of the higher-level theories for bulks, and at least
30\% less than the lowest reasonable theoretical prediction.

\subsection{Near-contact force}
\label{Sec:C33}

Putting aside the distinction between binding and cleavage energies,
and taking the model in the previous work\cite{Liu2012}
\emph{prima facie} we note
another difficulty that warrants further
analysis. The potential model employed is derived directly
from the two-parameter LJ potential of Ref.~\onlinecite{Girifalco2002}.
It gives the energy $\PhiLJ$ as a function of the distance parameter
$\nu$ describing the stretching of a single interplanar
distance during the cleavage process, while leaving other
interplanar distances unchanged.
\changed{For this paper we rewrite the energy in terms of
\[
x \equiv D-D_0 \equiv \nu-\nu_0,
\]
the difference
of the distance $D$ between the surfaces after cleavage,
and the graphite equilibrium interlayer spacing $D_0$
such that $\frac{d}{d x}\PhiLJ(0)=0$
($\nu_0$ is defined as the value of $\nu$ that minimises $\PhiLJ$).

Equation A6 of \Liue's supplementary material gives, after some algebra,
\begin{align}
\PhiLJ(x):=&\sum_{n=1}^{\infty} n\Phi_0(x+\nu_0+n\sigma),
\label{eqn:Phi}
\end{align}
where
\begin{align}
\Phi_0(\delta)=&\alpha \ECl\sigma^2
\lbrs \frac25(\sigma/\delta)^{10}-(\sigma/\delta)^4 \rbrs
\label{eqn:Phi0}
\end{align}
is the interlayer potential between two layers and is summed over all
possible layer interactions to form $\PhiLJ$. Here
$\ECl$ is the cleavage energy (called the ``binding energy'' in \Liue)
while $\alpha$ and $\sigma$ are constants, the latter being related
to the interlayer spacing and taking the value $\sigma=0.3415$nm
in the previous analysis. By definition of $x$, the energy takes
its minimum value
at $x=0$ which gives $\nu_0=-0.0045$nm and $D_0=0.337$nm, in
agreement with experiment. The constant
$\alpha=10.676$nm\sps{-2} is chosen to make $\PhiLJ(0)=-\ECl$
and can be derived from A2-A6 of the supplementary material
for \Liue.

The potential $\PhiLJ$ can be considered a non-linear
extension to Hooke's law between planes, reducing to Hooke's law for
small displacements. When two graphite surfaces are parallel and
near equilibrium lattice spacing, the resulting force should be
proportional to the displacement, and should correctly reproduce
the $C_{33}$ coefficient of bulk graphite measured in previous
experiments\cite{Blakslee1970,Gauster1974,Wada1980,Bosak2007}.
In the model \eqref{eqn:Phi}, small displacements
occur when $x\equiv D-D_0\aeq 0$ such that
\begin{align}
\PhiLJ(x)\aeq& -\ECl + \frac{\tilde{C}_{33}}{2D_0}x^2,
&
F(x)\aeq& \tilde{C}_{33}\frac{x}{D_0}
\label{eqn:Fx}
\end{align}
where $F(x)$ is the cleavage force near `contact'}. Here
\begin{align}
\tilde{C}_{33}=&D_0\frac{d^2 \PhiLJ(x)}{d x^2}\big|_{x=0}
\label{eqn:C33}
\end{align}
is, in fact the elastic coefficient via cleavage, and cannot trivially
be compared with existing experiments on bulk graphite.
As discussed in Appendix~\ref{app:Elas}
we can correct \eqref{eqn:C33} to obtain the true $C_{33}$ coefficient
for stretching of bulk graphite via
\begin{align}
C_{33}=&\tilde{C}_{33} + \Delta C_{33},
\label{eqn:C33t}
\end{align}
where $\Delta C_{33}$ is a small correction.

It is thus clear that we can use $\PhiLJ(x)$ to evaluate the
interlayer elastic coefficient $C_{33}$ via \eqref{eqn:C33}
and \eqref{eqn:C33t}. Using $\ECl=0.19$\Jmm\ and
$D_0=0.337$nm we find $\tilde{C}_{33}=19.7$GPa
and $C_{33}=20.0$GPa (the two coefficients are close enough to
be used almost interchangeably), well outside the range
of previous experiments\cite{Blakslee1970,Gauster1974,Wada1980,Bosak2007}
which give values between 36.5GPa and 40.7GPa.

This causes a discrepancy in the overall physical and mechanical model
in \Liue. For all layers except the open surface, an interplane
elastic coefficient of $C_{33}=36.5$GPa [from the stress-strain
relationship (A1) presented in the supplementary material]
was used. However, at the surface $C_{33}$ is proportional to $\ECl$
for the potential $\PhiLJ(x)$ employed, and gives an elastic constant
$C_{33}=20.0$GPa [from the interplane potential
model $\PhiLJ(x)$ defined via \eqref{eqn:Phi}].
The experimental elastic constant can be matched by setting the
cleavage energy to $0.36\pm 0.02$\Jmm, however this is inconsistent
with the value $0.19$\Jmm\ deduced by other means in \Liue.

\subsection{Intermediate and long distance force}
\label{Sec:Stiffness}

\begin{figure}
\caption{AFM measurements of one of the cantilevers showing the effect
of in-plane stiffness on the profile. Dotted lines are provided as a
visual guide. The experimental setup is shown below, showing
the scanning direction in (e).\label{fig:Stiffness}}
\includegraphics[clip,width=1.0\linewidth]{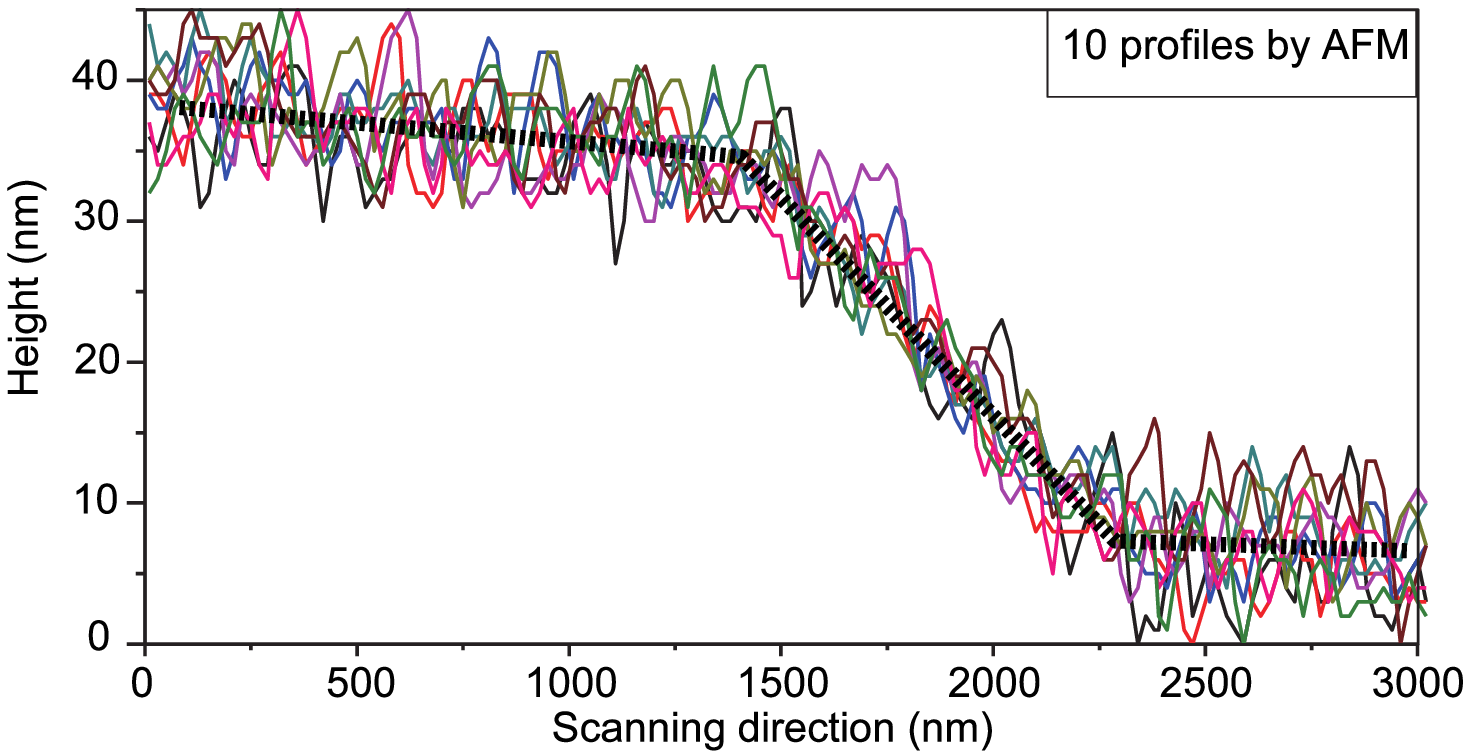}
\includegraphics[clip,width=1.0\linewidth]{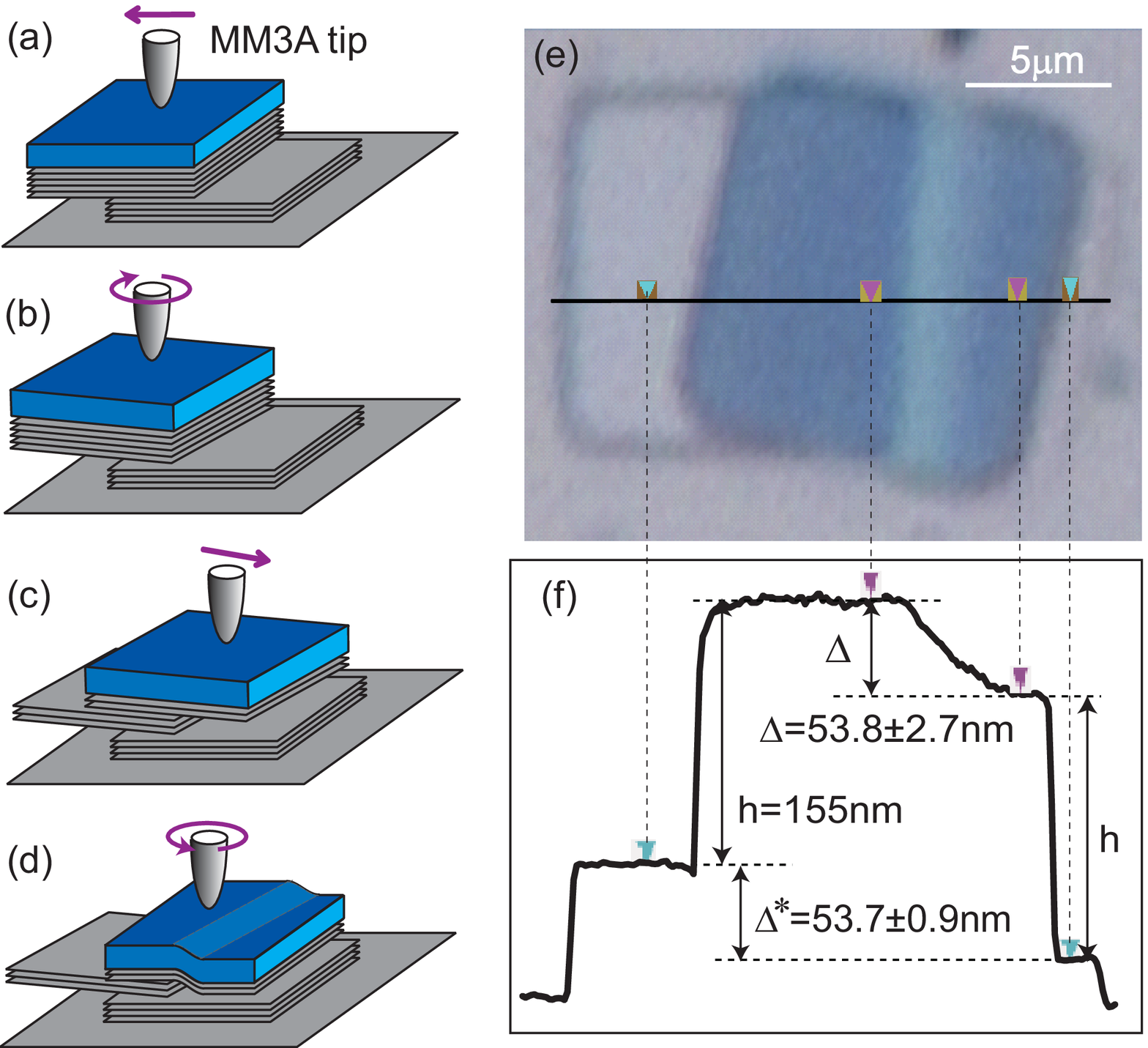}
\end{figure}
In the experimental setup reported in \Liue, graphite was arranged with
its layers lying parallel to the base of the
cantilevered segment.
The bending stiffness of the graphite cantilever is proportional to
the in-plane elastic coefficient of graphite (which is one of the
largest known at $\sim 1000$GPa) and the cube of the beam
thickness (e.g., $\sim 200$nm).
The strong bending stiffness makes bending deformation along the
cantilever energetically unfavourable.
This stiffness can be seen in the almost straight
line of the cantilever profile away from the base and pivot point,
as shown here in Figure~\ref{fig:Stiffness},
and in Figure S5 (the FEA illustration) and
Figure S7 (the AFM plots) in the same supplementary document.
It is clear that the cantilever (bending) deforms minimally until it
approaches contact with the base where the forces are sufficient
to overcome the internal stiffness.

As such the \emph{intermediate} part of the binding force will
be emphasised in any analysis of the experimental data, while the
\emph{long distance} binding will be harder to determine. This makes
evaluation of the binding energy difficult as it depends on
the integral of the force from infinite separation to finite
separation. This also means that the best fit [using \eqref{eqn:Phi}]
to a given experiment must match both the difficult to measure
longer-ranged forces, and the near and intermediate forces.
With only one free parameter matching all three ranges is
very difficult, and the intermediate region will likely
dominate the fit.

The experimental difficulties are further amplified by
the fact that theoretical understanding of the longer-ranged
van der Waals potential of graphite cleavage is limited. Here
high-level theory results\cite{Gould2009} are valid only for the
\emph{extreme} outer limit where corrections for finite
Dirac cones\cite{Gould2013-Cones} and other non-asymptotic physics
are not required.

\changed{This lack of theoretical insight comes from difficulties
in understanding the contribution of the $\pi_z$ and
$sp_2$ orbitals on the dispersion processes involved in cleavage.
While these dispersion forces contribute a signficant amount
to the intermediate region energy profile
($\sim 45$\% of the energy at contact for graphite, bigraphene
and exfoliation\cite{Gould2013-Model}),
they contribute comparitively less to the intermediate force.
Thus a failure to properly account for dispersion is less
problematic for interpretation of peak force measurements
compared to the cleavage energy.

Furthermore, the dispersion contribution is itself split into
two components, which dominate at different length scales.
Here the $sp_2$ orbitals dominate the dispersion in the
intermediate region, while $\pi_z$ orbitals dominate in
the asymptotic region.
The different force and length scales involved in dispersion
thus make extrapolation from the
near-contact region to the asymptotic region very difficult.
This makes mathematical fitting
of an all-region force unreliable without firm theoretical backing,
with consequences for the energy. It does not, however, affect
measurement of near-contact forces, which depend minimally on
the unknown asymptotics.}

\section{Alternate models}

As mentioned in Section~\ref{Sec:C33}, the Lennard-Jones
model \eqref{eqn:Phi} with
cleavage energy determined by best fit to experiment
failed to reproduce the known near-contact force of graphite.
Ideally any model potential should reproduce the correct
near-contact force by construction, via the experimentally measured
interlayer distance $D_0=0.334$nm, and the experimental
elastic coefficient $C_{33}=36.5$GPa. Additionally
the potential should decay\cite{Gould2009} as the van der Waals
power law $-C_2D^{-2}$ well away from contact.
A model potential can then be formed that allows the binding
energy to vary as a parameter, while matching these known
properties of graphite.
The LJ potential, with just two parameters, is not
able to simultaneously fit $D_0$, $C_{33}$ and $\ECl$, and leads to
$C_{33}\propto \ECl$. We thus propose two new models of the binding
energy that decouple the elastic coefficient from the cleavage energy,
and thus allow better reproduction of known properties of graphite.

Our first model is designed to reproduce the RPA energy curve in the
near-contact region when the input
energy takes the RPA value $\ECl=0.29$\Jmm.
Here, in addition to matching the binding distance and
elastic coefficient, we match 
the non-linear coefficient $C_{333}=-540$GPa obtained from
the RPA so that for small
$x/D_0$ the force obeys $F(x)\aeq C_{33}(x/D_0)+\half C_{333}(x/D_0)^2$.
One such ``near-contact model'' (NCM) is
\begin{align}
\PhiNCM(x)=&\frac{-\ECl}{1+c_2x^2+c_3x^3e^{-kx}}
\label{eqn:PhiNCM}
\end{align}
where $x=D-D_0$ is the deviation from the experimental lattice spacing.
Here $c_2=C_{33}/(2D_0\ECl)$ and $c_3=C_{333}/(6D_0^2\ECl)$
ensure that the first three derivatives of $\PhiNCM$ are
equal to their RPA values at contact $D=D_0$ for
arbitrary $\ECl$. Free parameter $k>0$ guarantees that
$\PhiNCM(x\to\infty)\propto x^{-2}$. We choose $k=8$ to ensure
that the force $\frac{\d\PhiNCM}{\d x}$ is single peaked for $x>0$,
and that the RPA data is approximately reproduced when
the RPA energy $\ECl=0.29$\Jmm\ is used.

Alternatively, one may wish to keep the LJ-like form of the model
potential, while reproducing the known physical properties
$D_0$ and $C_{33}$ as well as the correct $-C_2/D^2$ asymptotic
form (if not its coefficient $C_2$). This suggests a model
potential of general form
\begin{align}
\PhiLJp(x)=&\frac{\ECl}{p-2}
\lbrs
\frac{2}{(1+x/D_0)^p}-\frac{p}{(1+x/D_0)^2}
\rbrs
\label{eqn:PhiLJp}
\end{align}
where the exponent $p=C_{33}D_0/(2\ECl)$ is chosen to ensure the
experimental $C_{33}$ elastic coefficient can be reproduced.
Such an approach is perhaps
physically less justified than the above in the near contact
region, but does not require an RPA third-order elastic coefficient
as additional input, and may be more accurate away from contact.

\section{Results}

Following the previous work\cite{Liu2012}, we use the
three different model
potentials $\PhiLJ$, $\PhiNCM$ and $\PhiLJp$ [from respectively
equations \eqref{eqn:Phi}, \eqref{eqn:PhiNCM} and \eqref{eqn:PhiLJp}]
in FEA models to determine a model-dependent cleavage energy $\ECl$.
This was defined to be the value (for each model) that caused the
elastically deformed curves determined via FEA to best match
the experimentally measured profiles.
The variation of the predicted binding energy was very great
even within a given example geometry, ranging
from $\ECl=0.13${\Jmm} for the near-contact model $\PhiNCM$, to
$\ECl=0.19${\Jmm} for the LJ model $\Phi$ and
$\ECl=0.27${\Jmm} for the modified LJ model $\PhiLJp$.
The modified LJ model best matches previous theoretical predictions
of the cleavage energy, but the error bar across the three
models is too signficant for any result to be considered reliable.
This is unsurprising as the cleavage energy depends partly
on the very weak dispersive forces at large distance,
which are difficult to determine accurately via
the current experimental setup\cite{Liu2012}. Additionally, the asymptotic
van der Waals behaviour for cleavage is unknown, and none of
the models can be guaranteed to hold true in the more distant
limit.

\begin{figure}
\caption{Cleavage force per unit area under the three different
models \changed{plotted against the inter-layer distance $D$ (where the
equilibrium interlayer spacing is $D_0=0.334$nm)}.
The RPA \emph{bulk} layer binding force is included for illustrative
purposes, although it is not in general the same as the cleavage force.
\label{fig:Forces}}
\includegraphics[clip,width=\linewidth]{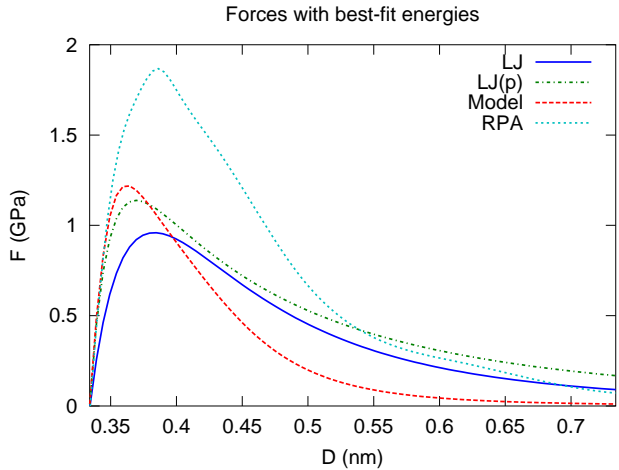}
\end{figure}
From the potential it is trivial to determine the force
$F=\frac{\d \Phi}{\d x}$, and in the
near-contact region this should be less model-dependent and given
more reliably by experiment.
Indeed by construction both $\PhiNCM$ and $\PhiLJp$
will predict identical linearly varying forces near the inter-layer
binding distance. Using the best fit $\ECl$ parameters for each model
yields distance-dependent forces displayed in Figure~\ref{fig:Forces}.
It is immediately apparent that while the force in the tail varies
greatly between the models, the position and magnitude of the ``peak''
force (the maximum) varies much less signficantly.
The ``peak force'' is a physically important property of
graphite, giving the force per unit area require to ``cleave''
graphite in two at a single surface. This agrees with the arguments
given in Section~\ref{Sec:Stiffness} that the analysis of
experiment of \Liue\ will be most sensitive to the
force in the intermediate range.

\begin{table}
\caption{Best fit cleavage energy, and corresponding peak
force and position for different models.
RPA results are provided for illustration, not comparison.
\label{tab:PeakForce}}
\begin{ruledtabular}
\begin{tabular}{lrrr}
Model & $\ECl$ (\Jmm) & $F_m$ (GPa) & $D_m$ (nm)
\\\hline
LJ $\Phi$ & 0.19 & 0.96 & 0.388
\\
NCM $\PhiNCM$ & 0.13 & 1.22 & 0.367
\\
LJ($p$) $\PhiLJp$ & 0.27 & 1.13 & 0.372
\\\hline
RPA Binding$^a$ & 0.29 & 1.87 & 0.385
\end{tabular}
$^a$ Using data from Ref.~\onlinecite{Lebegue2010} and additional data points.
\end{ruledtabular}
\end{table}
Using the best fit $\ECl$ for each of the three models,
we semi-analytically determine the
peak force $F_m$ and the inter-layer distance $D_m$ at which it occurs.
We tabulate the results in
Table~\ref{tab:PeakForce}, including the peak bulk layer binding
(as opposed to cleavage)
force predicted by the RPA for illustrative comparison.
The peak force varies by up to 27\% over all three models, which is
very good agreement for such different models. For the two
models $\PhiNCM$ and $\PhiLJp$ 
with the experimental elastic coefficient of bulk graphite
included as a parameter
the variation is even smaller still, with only a 5pm
difference to the position and an 8\% difference in the peak force.
This small variation is a very positive feature, as the
cleavage energy parameter varies by over 100\%.

The peak force predicted by the RPA is significantly larger
than those found via the analysis of the experiment
with any of the models. While at first glance this is worrying,
we note that the RPA data is for
the different problem of bulk layer binding (under uniaxial stretching
perpendicular to the planes) and cannot be directly compared.
Indeed one expects the force laws for cleavage, exfoliation and
bulk layer binding to show greater variation than the `binding'
energies as each obeys a different asymptotic
power law\cite{Gould2009}, with bulk Coulomb screening reducing
the dispersion force in cleavage and exfoliation.
This screening effect suggests that the peak force of cleavage
should be less than that of bulk layer binding, and thus one
should expect an apparent discrepancy between i) the
values of peak forces deduced from the experiments using the
three models, and ii) theoretical predictions from the RPA.
Full resolution of the differences between peak force
in bulk-layer binding and cleavage would require further
experiments and/or further RPA calculations (or other high-level
theory) for the cleavage geometry. These are beyond current
capabilities, however.

\section{Conclusion}

Overall we conclude that the binding energy evaluated through
the theoretical analysis of experimental results in the previous
work (\Liue) is likely to be
substantially underestimated. Firstly, the model employed in
\Liue\ yields a near-contact force law with an elastic
coefficient of $C_{33}=20.0$GPa, just over half that of previous reliable
experiments\cite{Blakslee1970,Gauster1974,Wada1980,Bosak2007}
with $C_{33}=38.6\pm 2.1$GPa.
The employed LJ model makes $C_{33}\propto \ECl$ suggesting that
$\ECl\aeq 0.36$\Jmm\ would be required to match the experiment
elastic parameter.
Secondly, the experimental setup and model potential
[see our \eqref{eqn:Phi} or (A6) of the supplementary material
of \Liue] are designed for the measurement of cleavage
energies $\ECl$, not the bulk-layer binding energies $\EB$ with which
their results were compared. It is believed that $\EB\aeq 0.85\ECl$
in graphitic systems, leading to a value $\EB\aeq 0.17$\Jmm\ from the
experimental analysis, approximately half the value estimated by high-level
theory\cite{Spanu2009,Podeszwa2010,Lebegue2010,Bjorkman2012}.

By attempting new fits to the same data, using model potentials
which include the experimental $C_{33}$ coefficient as an input,
we find that the predicted cleavage energy $\ECl$ varies dramatically
(by 100\%) depending on the model potential used.
However, the size and position of
the peak force deduced from the experiment
(the minimum force per unit area required to fully
``cleave'' graphite at a surface) is much less dependent on the model,
showing only a 27\% variation in magnitude across all three models,
reduced to 8\% between the two models that reproduce the experimental
$C_{33}$ coefficient. We thus conclude that the peak force per unit
area can be accurately predicted by this experiment, and that it is
$1.1\pm 0.15$GPa and occurs at an inter-layer spacing of
$0.377\pm 0.013$nm, where the error bar includes predicted
experimental and analysis errors, as well as variation
across the three interaction models.

The present experimental arrangement, and lack of asymptotic
van der Waals theory for cleavage makes it difficult to extract
accurate cleavage energies. However, similar experiments
(e.g., a graphite flake cantilever with fewer graphene layers),
in principle, can be devised that allow accurate measurement of the
force over a wider range of distances. Such experiments would enable
the cleavage energy to be determined experimentally by integration
over the force, and these will be the focus of future efforts.
Similarly it would be desirable to better understand at the
theoretical level the asymptotic van der Waals attractive
potential in graphite under cleavage,
and this is also a topic for future investigation.

\changed{Finally, we note that these difficulties in resolving
asymptotic properties
(and thus integrated quantities like the binding energy) are
likely to occur in analysis of indirect force measurements for
all layered materials, not just graphite.
Although they are expected to be most severe in
graphite due to the contribution of the $\pi_z$ dispersion.
We thus recommend against the use of the Lennard-Jones models
for layered systems, and recommend instead the use of something
like the LJ$(p)$ model \eqref{eqn:PhiLJp} when modelling
interlayer forces in systems with known elastic coefficients.}

\acknowledgments

We would like to thank Torbj{\"o}rn Bj{\"o}rkman for helpful discussion
on cleavage, exfoliation and binding in layered solids.
T.G. and J.F.D. were supported by ARC Discovery Grant DP1096240.
S.L. acknowledges financial support from the Universit\'e de Lorraine
through the program
``Soutien \`a la dimension internationale de la recherche''.
Q.S.Z. acknowledges
the financial support from NSFC through Grants No. 10832005 and
No. 10972113, the 973 Program through Grants No. 2007CB936803 and
No. 2013CB934200. J.Z.L. acknowledges Seed Grant 2013 from the
engineering faculty of Monash University.

\appendix
\section{Elastic coefficients}
\label{app:Elas}

In general the $C_{33}$ elastic coefficient is the ratio of interlayer
force to interlayer displacement induced by stretching all layers
evenly (with appropriate scaling).
By contrast the model used in \Liue\ involves displacement between
two particular layers (cleavage) with other layer spacing
kept essentially fixed, a process associated with elastic
constant $\tilde{C}_{33}$.
Employing the interlayer potential $\Phi_0$ [equation \eqref{eqn:Phi0}]
we can find the coefficient in the two different models
via [using $x=\nu-\nu_0$ in equation \eqref{eqn:C33} for
$\tilde{C}_{33}$ and a similar derivation for $C_{33}$]
\begin{align}
\tilde{C}_{33}=&D_0\frac{d^2}{d \nu^2}
\sum_{n=1}^{\infty} n\Phi_0(n\sigma + \nu)\big|_{\nu=\nu_0},
\\
C_{33}=&D_0\frac{d^2}{d \nu^2}
\sum_{n=1}^{\infty}\Phi_0(n\sigma + n\nu)\big|_{\nu=\nu_0},
\label{eqn:C33T}
\end{align}
where \eqref{eqn:C33T} should reproduce the true $C_{33}$ coefficient
of graphite for a good interlayer potential model.
The difference between these two coefficients is
\begin{align}
\Delta C_{33}\equiv&C_{33}-\tilde{C}_{33}
\nonumber\\
=&D_0\sum_{n=2}^{\infty}\frac{d^2}{d \nu^2}
\lbrs \Phi_0(n\sigma + n\nu) - n\Phi_0(n\sigma + \nu) \rbrs_{\nu=\nu_0}
\end{align}
since the $n=1$ terms cancel. We thus expect $\Delta C_{33}$
to be small compared to $C_{33}$ as $\Phi_0(2\sigma)\ll\Phi_0(\sigma)$
and $|\nu_0|\ll\sigma$.
Inserting the experimental parameters $\nu_0/\sigma=-0.13$,
$D_0=0.337$nm and $\EB=0.19$\Jmm\ gives
$\Delta C_{33}=0.26$GPa
and $\tilde{C}_{33}=19.7$GPa. As expected the correction is small at
around $1.3\%$.


\begin{thebibliography}{31}%
\makeatletter
\providecommand \@ifxundefined [1]{%
 \@ifx{#1\undefined}
}%
\providecommand \@ifnum [1]{%
 \ifnum #1\expandafter \@firstoftwo
 \else \expandafter \@secondoftwo
 \fi
}%
\providecommand \@ifx [1]{%
 \ifx #1\expandafter \@firstoftwo
 \else \expandafter \@secondoftwo
 \fi
}%
\providecommand \natexlab [1]{#1}%
\providecommand \enquote  [1]{``#1''}%
\providecommand \bibnamefont  [1]{#1}%
\providecommand \bibfnamefont [1]{#1}%
\providecommand \citenamefont [1]{#1}%
\providecommand \href@noop [0]{\@secondoftwo}%
\providecommand \href [0]{\begingroup \@sanitize@url \@href}%
\providecommand \@href[1]{\@@startlink{#1}\@@href}%
\providecommand \@@href[1]{\endgroup#1\@@endlink}%
\providecommand \@sanitize@url [0]{\catcode `\\12\catcode `\$12\catcode
  `\&12\catcode `\#12\catcode `\^12\catcode `\_12\catcode `\%12\relax}%
\providecommand \@@startlink[1]{}%
\providecommand \@@endlink[0]{}%
\providecommand \url  [0]{\begingroup\@sanitize@url \@url }%
\providecommand \@url [1]{\endgroup\@href {#1}{\urlprefix }}%
\providecommand \urlprefix  [0]{URL }%
\providecommand \Eprint [0]{\href }%
\providecommand \doibase [0]{http://dx.doi.org/}%
\providecommand \selectlanguage [0]{\@gobble}%
\providecommand \bibinfo  [0]{\@secondoftwo}%
\providecommand \bibfield  [0]{\@secondoftwo}%
\providecommand \translation [1]{[#1]}%
\providecommand \BibitemOpen [0]{}%
\providecommand \bibitemStop [0]{}%
\providecommand \bibitemNoStop [0]{.\EOS\space}%
\providecommand \EOS [0]{\spacefactor3000\relax}%
\providecommand \BibitemShut  [1]{\csname bibitem#1\endcsname}%
\let\auto@bib@innerbib\@empty
\bibitem [{\citenamefont {Novoselov}\ \emph {et~al.}(2004)\citenamefont
  {Novoselov}, \citenamefont {Geim}, \citenamefont {Morozov}, \citenamefont
  {Jiang}, \citenamefont {Zhang}, \citenamefont {Dubonos}, \citenamefont
  {Grigorieva},\ and\ \citenamefont {Firsov}}]{electric}%
  \BibitemOpen
  \bibfield  {author} {\bibinfo {author} {\bibfnamefont {K.}~\bibnamefont
  {Novoselov}}, \bibinfo {author} {\bibfnamefont {A.}~\bibnamefont {Geim}},
  \bibinfo {author} {\bibfnamefont {S.}~\bibnamefont {Morozov}}, \bibinfo
  {author} {\bibfnamefont {D.}~\bibnamefont {Jiang}}, \bibinfo {author}
  {\bibfnamefont {Y.}~\bibnamefont {Zhang}}, \bibinfo {author} {\bibfnamefont
  {S.}~\bibnamefont {Dubonos}}, \bibinfo {author} {\bibfnamefont
  {I.}~\bibnamefont {Grigorieva}}, \ and\ \bibinfo {author} {\bibfnamefont
  {A.}~\bibnamefont {Firsov}},\ }\href@noop {} {\bibfield  {journal} {\bibinfo
  {journal} {Science}\ }\textbf {\bibinfo {volume} {306}},\ \bibinfo {pages}
  {666} (\bibinfo {year} {2004})}\BibitemShut {NoStop}%
\bibitem [{\citenamefont {Geim}\ and\ \citenamefont {Novoselov}(2007)}]{r1}%
  \BibitemOpen
  \bibfield  {author} {\bibinfo {author} {\bibfnamefont {A.~K.}\ \bibnamefont
  {Geim}}\ and\ \bibinfo {author} {\bibfnamefont {K.~S.}\ \bibnamefont
  {Novoselov}},\ }\href@noop {} {\bibfield  {journal} {\bibinfo  {journal}
  {Nature Materials}\ }\textbf {\bibinfo {volume} {6}},\ \bibinfo {pages} {183}
  (\bibinfo {year} {2007})}\BibitemShut {NoStop}%
\bibitem [{\citenamefont {Katsnelson}(2006)}]{r2}%
  \BibitemOpen
  \bibfield  {author} {\bibinfo {author} {\bibfnamefont {M.~I.}\ \bibnamefont
  {Katsnelson}},\ }\href@noop {} {\bibfield  {journal} {\bibinfo  {journal}
  {Materials Today}\ }\textbf {\bibinfo {volume} {10}},\ \bibinfo {pages} {20}
  (\bibinfo {year} {2006})}\BibitemShut {NoStop}%
\bibitem [{\citenamefont {Castro~Neto}\ \emph {et~al.}(2009)\citenamefont
  {Castro~Neto}, \citenamefont {Guinea}, \citenamefont {Peres}, \citenamefont
  {Novoselov},\ and\ \citenamefont {Geim}}]{r3}%
  \BibitemOpen
  \bibfield  {author} {\bibinfo {author} {\bibfnamefont {A.~H.}\ \bibnamefont
  {Castro~Neto}}, \bibinfo {author} {\bibfnamefont {F.}~\bibnamefont {Guinea}},
  \bibinfo {author} {\bibfnamefont {N.~M.~R.}\ \bibnamefont {Peres}}, \bibinfo
  {author} {\bibfnamefont {K.~S.}\ \bibnamefont {Novoselov}}, \ and\ \bibinfo
  {author} {\bibfnamefont {A.~K.}\ \bibnamefont {Geim}},\ }\href@noop {}
  {\bibfield  {journal} {\bibinfo  {journal} {Rev. Mod. Phys.}\ }\textbf
  {\bibinfo {volume} {81}},\ \bibinfo {pages} {109} (\bibinfo {year}
  {2009})}\BibitemShut {NoStop}%
\bibitem [{\citenamefont {Katsnelson}, \citenamefont {Novoselov},\ and\
  \citenamefont {Geim}(2006)}]{misha1}%
  \BibitemOpen
  \bibfield  {author} {\bibinfo {author} {\bibfnamefont {M.~I.}\ \bibnamefont
  {Katsnelson}}, \bibinfo {author} {\bibfnamefont {K.~S.}\ \bibnamefont
  {Novoselov}}, \ and\ \bibinfo {author} {\bibfnamefont {A.~K.}\ \bibnamefont
  {Geim}},\ }\href@noop {} {\bibfield  {journal} {\bibinfo  {journal} {Nature
  Physics}\ }\textbf {\bibinfo {volume} {2}},\ \bibinfo {pages} {620} (\bibinfo
  {year} {2006})}\BibitemShut {NoStop}%
\bibitem [{\citenamefont {Hu}\ and\ \citenamefont {Gruner}(2013)}]{SamPat}%
  \BibitemOpen
  \bibfield  {author} {\bibinfo {author} {\bibfnamefont {L.}~\bibnamefont
  {Hu}}\ and\ \bibinfo {author} {\bibfnamefont {G.}~\bibnamefont {Gruner}},\
  }\href@noop {} {}\bibinfo {howpublished} {{U.S. Patent No. 8390589 B2}}
  (\bibinfo {year} {5 Mar 2013})\BibitemShut {NoStop}%
\bibitem [{\citenamefont {Kim}\ \emph {et~al.}(2009)\citenamefont {Kim},
  \citenamefont {Zhao}, \citenamefont {Jang}, \citenamefont {Lee},
  \citenamefont {Kim}, \citenamefont {Kim}, \citenamefont {Ahn}, \citenamefont
  {Kim}, \citenamefont {Choi},\ and\ \citenamefont {Hong}}]{kim}%
  \BibitemOpen
  \bibfield  {author} {\bibinfo {author} {\bibfnamefont {K.~S.}\ \bibnamefont
  {Kim}}, \bibinfo {author} {\bibfnamefont {Y.}~\bibnamefont {Zhao}}, \bibinfo
  {author} {\bibfnamefont {H.}~\bibnamefont {Jang}}, \bibinfo {author}
  {\bibfnamefont {S.~Y.}\ \bibnamefont {Lee}}, \bibinfo {author} {\bibfnamefont
  {J.~M.}\ \bibnamefont {Kim}}, \bibinfo {author} {\bibfnamefont {K.~S.}\
  \bibnamefont {Kim}}, \bibinfo {author} {\bibfnamefont {J.~H.}\ \bibnamefont
  {Ahn}}, \bibinfo {author} {\bibfnamefont {P.}~\bibnamefont {Kim}}, \bibinfo
  {author} {\bibfnamefont {J.}~\bibnamefont {Choi}}, \ and\ \bibinfo {author}
  {\bibfnamefont {B.~H.}\ \bibnamefont {Hong}},\ }\href@noop {} {\bibfield
  {journal} {\bibinfo  {journal} {Nature}\ }\textbf {\bibinfo {volume} {457}},\
  \bibinfo {pages} {706} (\bibinfo {year} {2009})}\BibitemShut {NoStop}%
\bibitem [{\citenamefont {Girifalco}\ and\ \citenamefont
  {Lad}(1956)}]{Girifalco1956}%
  \BibitemOpen
  \bibfield  {author} {\bibinfo {author} {\bibfnamefont {L.~A.}\ \bibnamefont
  {Girifalco}}\ and\ \bibinfo {author} {\bibfnamefont {R.~A.}\ \bibnamefont
  {Lad}},\ }\href {http://link.aip.org/link/?JCP/25/693/1} {\bibfield
  {journal} {\bibinfo  {journal} {The Journal of Chemical Physics}\ }\textbf
  {\bibinfo {volume} {25}},\ \bibinfo {pages} {693} (\bibinfo {year}
  {1956})}\BibitemShut {NoStop}%
\bibitem [{\citenamefont {Benedict}\ \emph {et~al.}(1998)\citenamefont
  {Benedict}, \citenamefont {Chopra}, \citenamefont {Cohen}, \citenamefont
  {Zettl}, \citenamefont {Louie},\ and\ \citenamefont {Crespi}}]{Benedict1998}%
  \BibitemOpen
  \bibfield  {author} {\bibinfo {author} {\bibfnamefont {L.~X.}\ \bibnamefont
  {Benedict}}, \bibinfo {author} {\bibfnamefont {N.~G.}\ \bibnamefont
  {Chopra}}, \bibinfo {author} {\bibfnamefont {M.~L.}\ \bibnamefont {Cohen}},
  \bibinfo {author} {\bibfnamefont {A.}~\bibnamefont {Zettl}}, \bibinfo
  {author} {\bibfnamefont {S.~G.}\ \bibnamefont {Louie}}, \ and\ \bibinfo
  {author} {\bibfnamefont {V.~H.}\ \bibnamefont {Crespi}},\ }\href@noop {}
  {\bibfield  {journal} {\bibinfo  {journal} {Chem. Phys. Letters}\ }\textbf
  {\bibinfo {volume} {286}},\ \bibinfo {pages} {490} (\bibinfo {year}
  {1998})}\BibitemShut {NoStop}%
\bibitem [{\citenamefont {Zacharia}, \citenamefont {Ulbricht},\ and\
  \citenamefont {Hertel}(2004)}]{Zacharia2004}%
  \BibitemOpen
  \bibfield  {author} {\bibinfo {author} {\bibfnamefont {R.}~\bibnamefont
  {Zacharia}}, \bibinfo {author} {\bibfnamefont {H.}~\bibnamefont {Ulbricht}},
  \ and\ \bibinfo {author} {\bibfnamefont {T.}~\bibnamefont {Hertel}},\
  }\href@noop {} {\bibfield  {journal} {\bibinfo  {journal} {Phys. Rev. B}\
  }\textbf {\bibinfo {volume} {69}},\ \bibinfo {pages} {155406} (\bibinfo
  {year} {2004})}\BibitemShut {NoStop}%
\bibitem [{\citenamefont {Dion}\ \emph {et~al.}(2004)\citenamefont {Dion},
  \citenamefont {Rydberg}, \citenamefont {Schr\"oder}, \citenamefont
  {Langreth},\ and\ \citenamefont {Lundqvist}}]{Dion2004}%
  \BibitemOpen
  \bibfield  {author} {\bibinfo {author} {\bibfnamefont {M.}~\bibnamefont
  {Dion}}, \bibinfo {author} {\bibfnamefont {H.}~\bibnamefont {Rydberg}},
  \bibinfo {author} {\bibfnamefont {E.}~\bibnamefont {Schr\"oder}}, \bibinfo
  {author} {\bibfnamefont {D.~C.}\ \bibnamefont {Langreth}}, \ and\ \bibinfo
  {author} {\bibfnamefont {B.~I.}\ \bibnamefont {Lundqvist}},\ }\href@noop {}
  {\bibfield  {journal} {\bibinfo  {journal} {Phys. Rev. Lett.}\ }\textbf
  {\bibinfo {volume} {92}},\ \bibinfo {pages} {246401} (\bibinfo {year}
  {2004})}\BibitemShut {NoStop}%
\bibitem [{\citenamefont {Chakarova-K\"ack}\ \emph {et~al.}(2006)\citenamefont
  {Chakarova-K\"ack}, \citenamefont {Schr\"oder}, \citenamefont {Lundqvist},\
  and\ \citenamefont {Langreth}}]{Chakarova2006}%
  \BibitemOpen
  \bibfield  {author} {\bibinfo {author} {\bibfnamefont {S.~D.}\ \bibnamefont
  {Chakarova-K\"ack}}, \bibinfo {author} {\bibfnamefont {E.}~\bibnamefont
  {Schr\"oder}}, \bibinfo {author} {\bibfnamefont {B.~I.}\ \bibnamefont
  {Lundqvist}}, \ and\ \bibinfo {author} {\bibfnamefont {D.~C.}\ \bibnamefont
  {Langreth}},\ }\href {http://link.aps.org/doi/10.1103/PhysRevLett.96.146107}
  {\bibfield  {journal} {\bibinfo  {journal} {Phys. Rev. Lett.}\ }\textbf
  {\bibinfo {volume} {96}},\ \bibinfo {pages} {146107} (\bibinfo {year}
  {2006})}\BibitemShut {NoStop}%
\bibitem [{\citenamefont {Ziambaras}\ \emph {et~al.}(2007)\citenamefont
  {Ziambaras}, \citenamefont {Kleis}, \citenamefont {Schr\"oder},\ and\
  \citenamefont {Hyldgaard}}]{Ziambaras2007}%
  \BibitemOpen
  \bibfield  {author} {\bibinfo {author} {\bibfnamefont {E.}~\bibnamefont
  {Ziambaras}}, \bibinfo {author} {\bibfnamefont {J.}~\bibnamefont {Kleis}},
  \bibinfo {author} {\bibfnamefont {E.}~\bibnamefont {Schr\"oder}}, \ and\
  \bibinfo {author} {\bibfnamefont {P.}~\bibnamefont {Hyldgaard}},\ }\href
  {http://link.aps.org/doi/10.1103/PhysRevB.76.155425} {\bibfield  {journal}
  {\bibinfo  {journal} {Phys. Rev. B}\ }\textbf {\bibinfo {volume} {76}},\
  \bibinfo {pages} {155425} (\bibinfo {year} {2007})}\BibitemShut {NoStop}%
\bibitem [{\citenamefont {Hasegawa}\ and\ \citenamefont
  {Nishidate}(2004)}]{Hasegawa2004}%
  \BibitemOpen
  \bibfield  {author} {\bibinfo {author} {\bibfnamefont {M.}~\bibnamefont
  {Hasegawa}}\ and\ \bibinfo {author} {\bibfnamefont {K.}~\bibnamefont
  {Nishidate}},\ }\href@noop {} {\bibfield  {journal} {\bibinfo  {journal}
  {Phys. Rev. B}\ }\textbf {\bibinfo {volume} {70}},\ \bibinfo {pages} {205431}
  (\bibinfo {year} {2004})}\BibitemShut {NoStop}%
\bibitem [{\citenamefont {Spanu}, \citenamefont {Sorella},\ and\ \citenamefont
  {Galli}(2009)}]{Spanu2009}%
  \BibitemOpen
  \bibfield  {author} {\bibinfo {author} {\bibfnamefont {L.}~\bibnamefont
  {Spanu}}, \bibinfo {author} {\bibfnamefont {S.}~\bibnamefont {Sorella}}, \
  and\ \bibinfo {author} {\bibfnamefont {G.}~\bibnamefont {Galli}},\ }\href
  {http://link.aps.org/doi/10.1103/PhysRevLett.103.196401} {\bibfield
  {journal} {\bibinfo  {journal} {Phys. Rev. Lett.}\ }\textbf {\bibinfo
  {volume} {103}},\ \bibinfo {pages} {196401} (\bibinfo {year}
  {2009})}\BibitemShut {NoStop}%
\bibitem [{\citenamefont {Leb\`egue}\ \emph {et~al.}(2010)\citenamefont
  {Leb\`egue}, \citenamefont {Harl}, \citenamefont {Gould}, \citenamefont
  {\'Angy\'an}, \citenamefont {Kresse},\ and\ \citenamefont
  {Dobson}}]{Lebegue2010}%
  \BibitemOpen
  \bibfield  {author} {\bibinfo {author} {\bibfnamefont {S.}~\bibnamefont
  {Leb\`egue}}, \bibinfo {author} {\bibfnamefont {J.}~\bibnamefont {Harl}},
  \bibinfo {author} {\bibfnamefont {T.}~\bibnamefont {Gould}}, \bibinfo
  {author} {\bibfnamefont {J.~G.}\ \bibnamefont {\'Angy\'an}}, \bibinfo
  {author} {\bibfnamefont {G.}~\bibnamefont {Kresse}}, \ and\ \bibinfo {author}
  {\bibfnamefont {J.~F.}\ \bibnamefont {Dobson}},\ }\href@noop {} {\bibfield
  {journal} {\bibinfo  {journal} {Phys. Rev. Lett.}\ }\textbf {\bibinfo
  {volume} {105}},\ \bibinfo {pages} {196401} (\bibinfo {year}
  {2010})}\BibitemShut {NoStop}%
\bibitem [{\citenamefont {Thrower}\ \emph {et~al.}(2013)\citenamefont
  {Thrower}, \citenamefont {Friis}, \citenamefont {Skov}, \citenamefont
  {Nilsson}, \citenamefont {Andersen}, \citenamefont {Ferrighi}, \citenamefont
  {Jørgensen}, \citenamefont {Baouche}, \citenamefont {Balog}, \citenamefont
  {Hammer},\ and\ \citenamefont {Hornekær}}]{Thrower2013}%
  \BibitemOpen
  \bibfield  {author} {\bibinfo {author} {\bibfnamefont {J.~D.}\ \bibnamefont
  {Thrower}}, \bibinfo {author} {\bibfnamefont {E.~E.}\ \bibnamefont {Friis}},
  \bibinfo {author} {\bibfnamefont {A.~L.}\ \bibnamefont {Skov}}, \bibinfo
  {author} {\bibfnamefont {L.}~\bibnamefont {Nilsson}}, \bibinfo {author}
  {\bibfnamefont {M.}~\bibnamefont {Andersen}}, \bibinfo {author}
  {\bibfnamefont {L.}~\bibnamefont {Ferrighi}}, \bibinfo {author}
  {\bibfnamefont {B.}~\bibnamefont {Jørgensen}}, \bibinfo {author}
  {\bibfnamefont {S.}~\bibnamefont {Baouche}}, \bibinfo {author} {\bibfnamefont
  {R.}~\bibnamefont {Balog}}, \bibinfo {author} {\bibfnamefont
  {B.}~\bibnamefont {Hammer}}, \ and\ \bibinfo {author} {\bibfnamefont
  {L.}~\bibnamefont {Hornekær}},\ }\href
  {http://pubs.acs.org/doi/abs/10.1021/jp404240h} {\bibfield  {journal}
  {\bibinfo  {journal} {The Journal of Physical Chemistry C}\ }\textbf
  {\bibinfo {volume} {117}},\ \bibinfo {pages} {13520} (\bibinfo {year}
  {2013})}\BibitemShut {NoStop}%
\bibitem [{\citenamefont {Bu\ifmmode~\check{c}\else \v{c}\fi{}ko}\ \emph
  {et~al.}(2013)\citenamefont {Bu\ifmmode~\check{c}\else \v{c}\fi{}ko},
  \citenamefont {Leb\`egue}, \citenamefont {Hafner},\ and\ \citenamefont
  {\'Angy\'an}}]{TSvasp}%
  \BibitemOpen
  \bibfield  {author} {\bibinfo {author} {\bibfnamefont {T.~c.~v.}\
  \bibnamefont {Bu\ifmmode~\check{c}\else \v{c}\fi{}ko}}, \bibinfo {author}
  {\bibfnamefont {S.}~\bibnamefont {Leb\`egue}}, \bibinfo {author}
  {\bibfnamefont {J.}~\bibnamefont {Hafner}}, \ and\ \bibinfo {author}
  {\bibfnamefont {J.~G.}\ \bibnamefont {\'Angy\'an}},\ }\href
  {http://link.aps.org/doi/10.1103/PhysRevB.87.064110} {\bibfield  {journal}
  {\bibinfo  {journal} {Phys. Rev. B}\ }\textbf {\bibinfo {volume} {87}},\
  \bibinfo {pages} {064110} (\bibinfo {year} {2013})}\BibitemShut {NoStop}%
\bibitem [{\citenamefont {Liu}\ \emph {et~al.}(2012)\citenamefont {Liu},
  \citenamefont {Liu}, \citenamefont {Cheng}, \citenamefont {Li}, \citenamefont
  {Wang},\ and\ \citenamefont {Zheng}}]{Liu2012}%
  \BibitemOpen
  \bibfield  {author} {\bibinfo {author} {\bibfnamefont {Z.}~\bibnamefont
  {Liu}}, \bibinfo {author} {\bibfnamefont {J.~Z.}\ \bibnamefont {Liu}},
  \bibinfo {author} {\bibfnamefont {Y.}~\bibnamefont {Cheng}}, \bibinfo
  {author} {\bibfnamefont {Z.}~\bibnamefont {Li}}, \bibinfo {author}
  {\bibfnamefont {L.}~\bibnamefont {Wang}}, \ and\ \bibinfo {author}
  {\bibfnamefont {Q.}~\bibnamefont {Zheng}},\ }\href
  {http://link.aps.org/doi/10.1103/PhysRevB.85.205418} {\bibfield  {journal}
  {\bibinfo  {journal} {Phys. Rev. B}\ }\textbf {\bibinfo {volume} {85}},\
  \bibinfo {pages} {205418} (\bibinfo {year} {2012})}\BibitemShut {NoStop}%
\bibitem [{\citenamefont {Bj\"orkman}\ \emph {et~al.}(2012)\citenamefont
  {Bj\"orkman}, \citenamefont {Gulans}, \citenamefont {Krasheninnikov},\ and\
  \citenamefont {Nieminen}}]{Bjorkman2012}%
  \BibitemOpen
  \bibfield  {author} {\bibinfo {author} {\bibfnamefont {T.}~\bibnamefont
  {Bj\"orkman}}, \bibinfo {author} {\bibfnamefont {A.}~\bibnamefont {Gulans}},
  \bibinfo {author} {\bibfnamefont {A.~V.}\ \bibnamefont {Krasheninnikov}}, \
  and\ \bibinfo {author} {\bibfnamefont {R.~M.}\ \bibnamefont {Nieminen}},\
  }\href {http://link.aps.org/doi/10.1103/PhysRevLett.108.235502} {\bibfield
  {journal} {\bibinfo  {journal} {Phys. Rev. Lett.}\ }\textbf {\bibinfo
  {volume} {108}},\ \bibinfo {pages} {235502} (\bibinfo {year}
  {2012})}\BibitemShut {NoStop}%
\bibitem [{\citenamefont {Gould}, \citenamefont {Simpkins},\ and\ \citenamefont
  {Dobson}(2008)}]{Gould2008}%
  \BibitemOpen
  \bibfield  {author} {\bibinfo {author} {\bibfnamefont {T.}~\bibnamefont
  {Gould}}, \bibinfo {author} {\bibfnamefont {K.}~\bibnamefont {Simpkins}}, \
  and\ \bibinfo {author} {\bibfnamefont {J.~F.}\ \bibnamefont {Dobson}},\
  }\href@noop {} {\bibfield  {journal} {\bibinfo  {journal} {Phys. Rev. B}\
  }\textbf {\bibinfo {volume} {77}},\ \bibinfo {pages} {165134} (\bibinfo
  {year} {2008})}\BibitemShut {NoStop}%
\bibitem [{\citenamefont {Podeszwa}(2010)}]{Podeszwa2010}%
  \BibitemOpen
  \bibfield  {author} {\bibinfo {author} {\bibfnamefont {R.}~\bibnamefont
  {Podeszwa}},\ }\href {http://link.aip.org/link/?JCP/132/044704/1} {\bibfield
  {journal} {\bibinfo  {journal} {The Journal of Chemical Physics}\ }\textbf
  {\bibinfo {volume} {132}},\ \bibinfo {pages} {044704} (\bibinfo {year}
  {2010})}\BibitemShut {NoStop}%
\bibitem [{\citenamefont {Bj\"orkman}\ and\ \citenamefont
  {Gulans}()}]{Bjorkman2012-PC}%
  \BibitemOpen
  \bibfield  {author} {\bibinfo {author} {\bibfnamefont {T.}~\bibnamefont
  {Bj\"orkman}}\ and\ \bibinfo {author} {\bibfnamefont {A.}~\bibnamefont
  {Gulans}},\ }\href@noop {} {}\bibinfo {note} {(private
  communication)}\BibitemShut {NoStop}%
\bibitem [{\citenamefont {Girifalco}\ and\ \citenamefont
  {Hodak}(2002)}]{Girifalco2002}%
  \BibitemOpen
  \bibfield  {author} {\bibinfo {author} {\bibfnamefont {L.~A.}\ \bibnamefont
  {Girifalco}}\ and\ \bibinfo {author} {\bibfnamefont {M.}~\bibnamefont
  {Hodak}},\ }\href@noop {} {\bibfield  {journal} {\bibinfo  {journal} {Phys.
  Rev. B}\ }\textbf {\bibinfo {volume} {65}},\ \bibinfo {pages} {125404}
  (\bibinfo {year} {2002})}\BibitemShut {NoStop}%
\bibitem [{\citenamefont {Blakslee}\ \emph {et~al.}(1970)\citenamefont
  {Blakslee}, \citenamefont {Proctor}, \citenamefont {Seldin}, \citenamefont
  {Spence},\ and\ \citenamefont {Weng}}]{Blakslee1970}%
  \BibitemOpen
  \bibfield  {author} {\bibinfo {author} {\bibfnamefont {O.~L.}\ \bibnamefont
  {Blakslee}}, \bibinfo {author} {\bibfnamefont {D.~G.}\ \bibnamefont
  {Proctor}}, \bibinfo {author} {\bibfnamefont {E.~J.}\ \bibnamefont {Seldin}},
  \bibinfo {author} {\bibfnamefont {G.~B.}\ \bibnamefont {Spence}}, \ and\
  \bibinfo {author} {\bibfnamefont {T.}~\bibnamefont {Weng}},\ }\href
  {http://link.aip.org/link/?JAP/41/3373/1} {\bibfield  {journal} {\bibinfo
  {journal} {Journal of Applied Physics}\ }\textbf {\bibinfo {volume} {41}},\
  \bibinfo {pages} {3373} (\bibinfo {year} {1970})}\BibitemShut {NoStop}%
\bibitem [{\citenamefont {Gauster}\ and\ \citenamefont
  {Fritz}(1974)}]{Gauster1974}%
  \BibitemOpen
  \bibfield  {author} {\bibinfo {author} {\bibfnamefont {W.~B.}\ \bibnamefont
  {Gauster}}\ and\ \bibinfo {author} {\bibfnamefont {I.~J.}\ \bibnamefont
  {Fritz}},\ }\href {http://link.aip.org/link/?JAP/45/3309/1} {\bibfield
  {journal} {\bibinfo  {journal} {Journal of Applied Physics}\ }\textbf
  {\bibinfo {volume} {45}},\ \bibinfo {pages} {3309} (\bibinfo {year}
  {1974})}\BibitemShut {NoStop}%
\bibitem [{\citenamefont {Wada}, \citenamefont {Clarke},\ and\ \citenamefont
  {Solin}(1980)}]{Wada1980}%
  \BibitemOpen
  \bibfield  {author} {\bibinfo {author} {\bibfnamefont {N.}~\bibnamefont
  {Wada}}, \bibinfo {author} {\bibfnamefont {R.}~\bibnamefont {Clarke}}, \ and\
  \bibinfo {author} {\bibfnamefont {S.}~\bibnamefont {Solin}},\ }\href
  {http://www.sciencedirect.com/science/article/pii/0038109880908728}
  {\bibfield  {journal} {\bibinfo  {journal} {Solid State Communications}\
  }\textbf {\bibinfo {volume} {35}},\ \bibinfo {pages} {675 } (\bibinfo {year}
  {1980})}\BibitemShut {NoStop}%
\bibitem [{\citenamefont {Bosak}\ \emph {et~al.}(2007)\citenamefont {Bosak},
  \citenamefont {Krisch}, \citenamefont {Mohr}, \citenamefont {Maultzsch},\
  and\ \citenamefont {Thomsen}}]{Bosak2007}%
  \BibitemOpen
  \bibfield  {author} {\bibinfo {author} {\bibfnamefont {A.}~\bibnamefont
  {Bosak}}, \bibinfo {author} {\bibfnamefont {M.}~\bibnamefont {Krisch}},
  \bibinfo {author} {\bibfnamefont {M.}~\bibnamefont {Mohr}}, \bibinfo {author}
  {\bibfnamefont {J.}~\bibnamefont {Maultzsch}}, \ and\ \bibinfo {author}
  {\bibfnamefont {C.}~\bibnamefont {Thomsen}},\ }\href
  {http://link.aps.org/doi/10.1103/PhysRevB.75.153408} {\bibfield  {journal}
  {\bibinfo  {journal} {Phys. Rev. B}\ }\textbf {\bibinfo {volume} {75}},\
  \bibinfo {pages} {153408} (\bibinfo {year} {2007})}\BibitemShut {NoStop}%
\bibitem [{\citenamefont {Gould}, \citenamefont {Gray},\ and\ \citenamefont
  {Dobson}(2009)}]{Gould2009}%
  \BibitemOpen
  \bibfield  {author} {\bibinfo {author} {\bibfnamefont {T.}~\bibnamefont
  {Gould}}, \bibinfo {author} {\bibfnamefont {E.}~\bibnamefont {Gray}}, \ and\
  \bibinfo {author} {\bibfnamefont {J.~F.}\ \bibnamefont {Dobson}},\
  }\href@noop {} {\bibfield  {journal} {\bibinfo  {journal} {Phys. Rev. B}\
  }\textbf {\bibinfo {volume} {79}},\ \bibinfo {pages} {113402} (\bibinfo
  {year} {2009})}\BibitemShut {NoStop}%
\bibitem [{\citenamefont {Gould}, \citenamefont {Dobson},\ and\ \citenamefont
  {Leb\`egue}(2013)}]{Gould2013-Cones}%
  \BibitemOpen
  \bibfield  {author} {\bibinfo {author} {\bibfnamefont {T.}~\bibnamefont
  {Gould}}, \bibinfo {author} {\bibfnamefont {J.~F.}\ \bibnamefont {Dobson}}, \
  and\ \bibinfo {author} {\bibfnamefont {S.}~\bibnamefont {Leb\`egue}},\ }\href
  {http://link.aps.org/doi/10.1103/PhysRevB.87.165422} {\bibfield  {journal}
  {\bibinfo  {journal} {Phys. Rev. B}\ }\textbf {\bibinfo {volume} {87}},\
  \bibinfo {pages} {165422} (\bibinfo {year} {2013})}\BibitemShut {NoStop}%
\bibitem [{\citenamefont {Gould}, \citenamefont {Leb\`egue},\ and\
  \citenamefont {Dobson}(2013)}]{Gould2013-Model}%
  \BibitemOpen
  \bibfield  {author} {\bibinfo {author} {\bibfnamefont {T.}~\bibnamefont
  {Gould}}, \bibinfo {author} {\bibfnamefont {S.}~\bibnamefont {Leb\`egue}}, \
  and\ \bibinfo {author} {\bibfnamefont {J.~F.}\ \bibnamefont {Dobson}},\
  }\href {http://stacks.iop.org/0953-8984/25/i=44/a=445010} {\bibfield
  {journal} {\bibinfo  {journal} {Journal of Physics: Condensed Matter}\
  }\textbf {\bibinfo {volume} {25}},\ \bibinfo {pages} {445010} (\bibinfo
  {year} {2013})}\BibitemShut {NoStop}%
\end{thebibliography}
%

\end{document}